# Anionic nanoparticle-lipid membrane interactions: the protonation of anionic ligands at the membrane surface reduces membrane disruption

Sebastian Salassi[a], Ester Canepa[b], Riccardo Ferrando[a] and Giulia Rossi [a†]

[a]Department of Physics, University of Genoa, Via Dodecaneso 33, 16146 Genoa, Italy
[b]Department of Chemistry and Industrial Chemistry, University of Genoa, Via Dodecaneso 31, 16146 Genoa, Italy
[†]rossig@fisica.unige.it

**Abstract**
Monolayer-protected gold nanoparticles (Au NPs) are promising biomedical tools with applications to diagnosis and therapy, thanks to their biocompatibility and versatility. Here we show how the NP surface functionalization can drive the mechanism of interaction with lipid membranes. In particular, we show that the spontaneous protonation of anionic carboxylic groups on the NP surface can make the NP-membrane interaction faster and less disruptive.

## Introduction

Inorganic nanoparticles (NPs), often functionalized by organic and biocompatible ligand shells, offer a number of opportunities in biomedicine. Imaging[1], photothermal therapies[2,3] and targeted drug delivery[4] are only a few of the applications involving ligand-protected inorganic NPs. Yet, the rational design of these inherently multivalent nanoagents remains a challenge[5]. This is due, on the one hand, to the difficulty of achieving simultaneous control of many different physico-chemical characteristics of the NP such as the size, shape, solubility and ligand functionality and, on the other hand, to the complexity of the NP interactions with the target biological environment.

Monolayer-protected Au nanoparticles (Au NPs) have emerged as a reference system in the field. Au is certainly of practical interest, as its optical properties can be exploited for *in vitro* sensing and *in vivo* imaging[6], delivery applications[7] and photothermal therapies[3], which have already entered clinical trials[8]. Moreover, as it is nowadays possible to achieve an excellent control of their composition and surface patterning[5], Au NPs are ideal to investigate the basic and general principles of their interactions with different biological targets.

Surface charge and the degree of hydrophilicity are important factors driving the fate of functionalized NPs inside the organism. Surface charge and hydrophilicity influence NP solubility and their circulation time in the blood stream; they affect the NP interactions with serum proteins and the stability of the protein corona[9–11]; eventually, they contribute to determine the NP interaction with the cell membrane[12,13]. For anionic NPs interacting with model zwitterionic lipid membranes, surface charges contribute to the interaction with a repulsive electrostatic term, while hydrophobicity drives the possible embedding of the NP in the membrane core[14–16].

Several computational studies have investigated the molecular mechanisms by which monolayer-protected, anionic Au NPs interact with zwitterionic lipid membranes[14,17,18]. The embedding of the NP into the membrane core is favorable from a thermodynamic point of view, but requires the overcoming of large energy barriers. The charged NP ligands need to translocate through the hydrophobic membrane core to anchor the NP to the membrane. The energetic cost of this transition has been estimated similar[17] or lower[18] than the cost of single monovalent ion translocations, depending on the arrangement of the ligands on the NP surface, on the type of lipid and on the force field used to perform the free energy calculation. Both *in silico* and experimental data support the idea that the presence of defects in lipid packing, such as those found at the edges of bicelles or supported lipid bilayers, may significantly reduce the cost of inserting the NP into the bilayer [19].

Here we consider zwitterionic membranes and Au NPs with a fixed size functionalized by a mixture of hydrophobic and hydrophilic ligands in a ratio of 1:1. The hydrophobic ligands are octane thiols (OT) and the hydrophilic ligands are anionic 11-mercaptoundecanoic acids (MUA). The atomistic structure of the ligands is shown in Fig. S1 of the ESI. These NP core and surface composition have become a reference for the study of NP-membrane interactions[14,17,20–25], and many experimental and computational results indicate the existence of a stable NP-membrane interaction.

We show, by a computational approach, that the NP-membrane interaction can be influenced by the protonation state of the charged ligands. Our calculations show that i) protonation of the carboxylate terminal group of the anionic ligands is more and more favorable as the NP approaches the membrane, ii) protonation may facilitate the NP-membrane interaction by lowering the free energy barriers along the pathway to the embedding of the NP in the membrane core and iii) the translocation of protonated (i.e., –COOH terminated) ligand makes the NP-membrane interaction a completely non-disruptive process, with little if no alteration of membrane integrity during the interaction process.

## Methods

The time scale of NP-membrane interaction is too long to be approached by unbiased molecular dynamics (MD) simulations with full atomistic resolution. Here, we rely on the use of the popular coarse-grained (CG) Martini force field[26]. As we are interested in the study of charged NPs and zwitterionic model lipid membranes, we adopt the polarizable water version (PW) of the Martini CG force field[27,28]. In this model, the CG water bead retains some orientational and deformational polarizability. The PW force field allows for a more accurate description of charge-charge interactions in non-polar environments, such as the membrane core, interactions which are severely underestimated in the non-polarizable version of the force field.

The PW model of the NP, as already described and validated in our earlier work[14,17], comprises an atomistic description of the Au core, with a diameter of 2 nm, with CG representation of the ligand shell. Hydrophobic OT ligands are described by a chain of two $C_1$ Martini beads while the charged MUA ligands are described by a chain of three hydrophobic $C_1$ beads and one terminal negatively charged $Q_{da}$ bead. The CG mapping is shown in Fig. S1. The NPs are covered by 30 OT ligands and 30 MUA ligands with a random grafting on the Au core. The model lipid membrane is composed by 512 zwitterionic POPC lipids. We simulated the NP-membrane interaction by means of unbiased MD and sampled the free energy landscape of the NP-membrane complex via metadynamics calculations[29]. Na counter ions were added to the solution to balance the NP charge. We performed simulations in the NPT ensemble, with the velocity–rescale thermostat[30] to set the temperature to 310 K. The pressure was kept constant to 1 bar with a semi-isotropic coupling using the Berendsen[31] and the Parrinello–Rahman[32] algorithms for the equilibration and production runs respectively. We used a timestep of 20 fs. More details on the unbiased MD set-up are reported in the ESI. Metadynamics simulations were run following the simulation setup described in our previous work[17] and recalled in the ESI. All simulations were performed with GROMACS 2016 and Plumed 2.3[33].

## Results and discussion

Our previous simulations show that the interaction of anionic Au NPs with zwitterionic lipid bilayers is a process that involves the transition between different metastable states[14]. One transition, in particular, determines the overall time scale of the interaction. We refer to it as to the "anchoring transition": one by one, the hydrophilic ligands of the NP, initially bound to the headgroup region of the entrance leaflet, cross the hydrophobic membrane core to bind to the distal leaflet. An example of anchoring transition is shown in the left panel of Fig. 1.

We previously showed, for NPs with an *ordered* arrangement of ligands on the surface, that the translocation of charged ligands can involve significant membrane deformations and transient membrane poration[14,17]. Here we repeated the procedure for a Au NP with a *random* arrangement of OT and MUA ligands on the surface. As the transition requires the charged ligands to overcome a significant free energy barrier, the process cannot be observed during unbiased MD runs. We thus use metadynamics to accelerate the process. As in our previous work[17], the dynamics of a single charged terminal bead of one ligand is biased along the reaction coordinate, which is the $z$ component of the distance between the terminal group of the biased ligand and the center of mass of the membrane, $d_z$. The visual inspection of the biased trajectories suggests that the charged ligand translocation induces significant membrane deformations, as shown in the left panel of Fig. 1. Moreover, in 6 out of 8 translocation processes we observed at least one water bead being transferred across the membrane together with the anchoring ligand. The right panel of Fig. 1 shows the average number of contacts between the charged terminal of the biased ligand and water as a function of the reaction coordinate: when the charged ligand terminals approach the center of the surface have been predicted to shift the pK$_a$ of similar ligands up to 7.6[34] (upper edge of the shaded area). Using 6.3 as reference pK$_a$ value in water[35], the light-blue profile is obtained and the error bars deriving from our metadynamics calculations lay within the shaded area. The shaded grey areas indicate the distribution of lipid heads and glycerol groups.

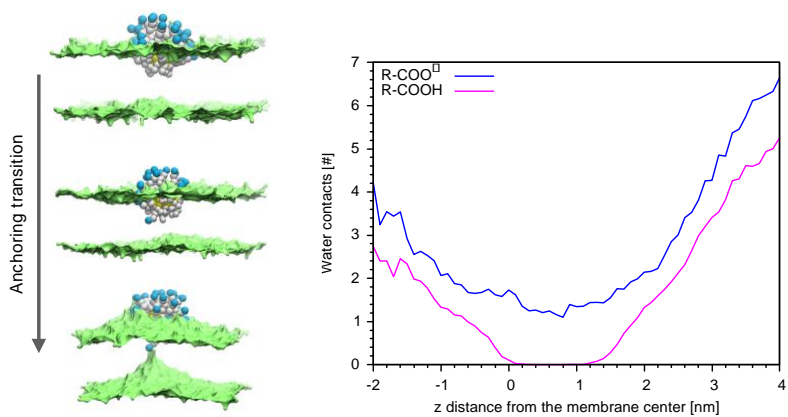

**Fig. 1** Left: hydrophobic ligand beads in white, anionic terminal groups in blue, lipid heads in green (surface representation), lipid tails and water not shown. From the top to the bottom, a charged ligand undergoes the anchoring transition. Right: the average number of contacts between the biased ligand terminal and CG water beads vs. of $d_z$.

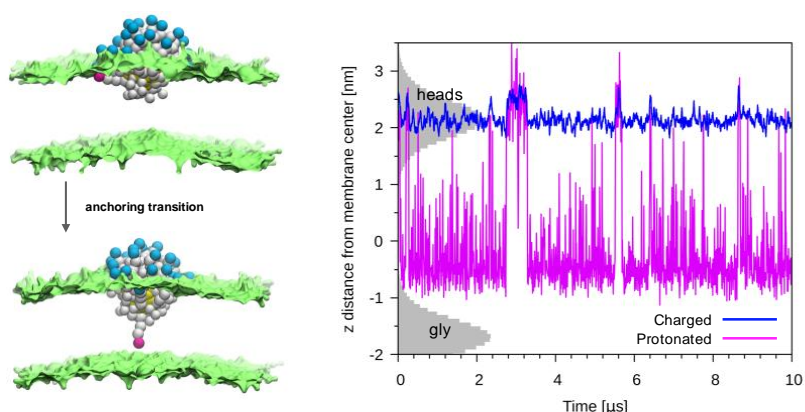

**Fig. 2** Left panel: A protonated ligand makes the anchoring transition, without deforming the lipid membrane. Color code as in Fig. 1, protonated $P_3$ bead in fuchsia. Right panel: Plot of the $d_z$ for a protonated ligand (fuchsia) or of a charged one (blue), during an unbiased simulation with a single protonated ligand. In the starting configuration both the protonated and the charged ligand were bound to the entrance leaflet (2 < z < 2.5 nm). When bound to the distal leaflet, the protonated ligand has $d_z$ in –1.0 < z < 0.5 nm range. The shaded grey areas indicate the distribution of lipid heads (top leaflet) and glycerol groups (bottom leaflet).

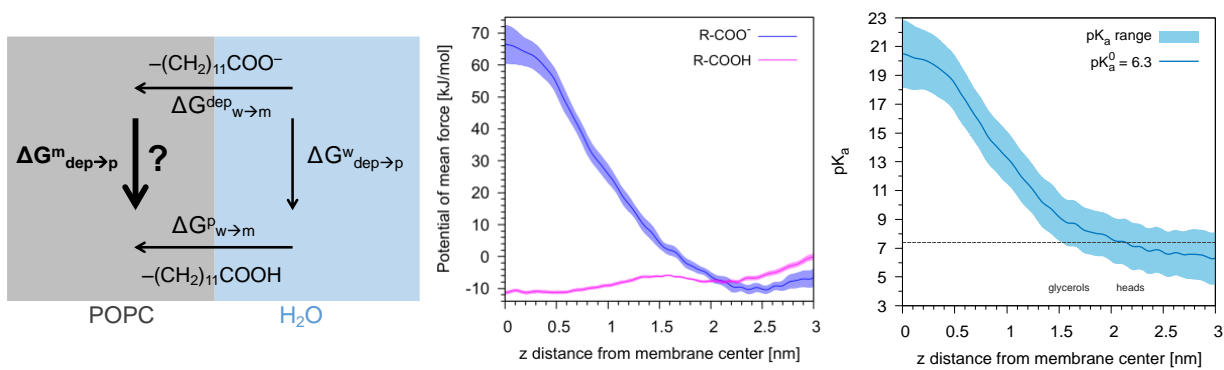

**Fig. 3** Left: the thermodynamic cycle used to calculate the $pK_a$ of the ligand vs. $d_z$. Center: the potential of mean force for the protonated and deprotonated ligand as a function of $d_z$. Shaded areas show the statistical error on ΔG, which was obtained by averaging over 8 metadynamics runs. Right: the $pK_a$ of the ligand vs. $d_z$. The shaded area shows a conservative estimate of the indeterminacy with which the $pK_a$ of the ligand in water is known: the free ligand has a $pK_a$ of 5 (bottom edge of the shaded area for $d_z$ = 3 nm), while cooperative effects on the NP

membrane they are still well hydrated (we remark that each CG water bead represents 4 water molecules).

MUA ligands have a pK$_a$ of 5 in water at physiological pH. When adsorbed on the surface of a NP, though, their pK$_a$ changes and shifts to larger values. Moglianetti *et al.*[35] have measured an average pK$_a$ of 6.3 for MUA ligands adsorbed on the surface of 4-5 nm Au NPs, suggesting that about one tenth of the MUA ligands are indeed protonated at physiological pH. The interaction with the membrane can induce changes of the protonation state, as well[36,37]. As the anionic ligands interacts with the lipid headgroups and with the membrane interior, they remain hydrated (Fig. 1) and thus in contact with a proton source. Could a change of the NP protonation state be responsible for a less disruptive character of NP-membrane interactions?

To answer this question, we first checked if and how the ligand translocation could be affected by protonation. We changed the Martini type of one charged ligand terminal to represent a protonated carboxyl. According to the Martini scheme, the new bead type is P$_3$, which is neutral but preserves a strong polar character and affinity to the lipid headgroup region.

We then performed an unbiased MD simulation starting form a configuration in which the protonated ligand was in contact with the entrance leaflet (top left panel of Fig. 2). In this condition we observe many spontaneous anchoring and detachment events of the protonated ligand to and from the distal leaflet. The fast anchoring kinetics, shown in the right panel of Fig. 2, indicates the presence of a much smaller anchoring barrier than for the charged ligand. Translocation events did not cause evident membrane deformations, as shown in left panel of Fig. 2, and no translocation of water beads was ever observed during the protonated ligand anchoring, as shown in the right panel of Fig. 1. The unbiased run thus suggests that, if the interaction of the NP with the membrane could induce protonation of the charged ligands, this would turn into a faster and less disruptive interaction with the membrane.

More quantitatively, we used metadynamics to calculate the free energy barriers for the translocation of one charged or one protonated ligand (bound to the NP, as in Fig. 1) across the membrane. It has been shown that the embedding of the NP into the membrane core happens via a sequence of single-ligand translocation events, all characterized by similar energy barriers[18]. The anchoring barrier of the negatively charged ligand is 76 ± 6 kJ/mol. The free energy profile of the protonated ligand shows a first small barrier of about 2 kJ/mol followed by a substantially flat landscape. The free energy difference between the two metastable states (in the entrance and distal leaflets) is about –4 kJ/mol, in favor of the anchored configuration. These small barriers and free energy differences between metastable states are consistent with the fast kinetics observed during the unbiased run.

We then aimed at the calculation of the effective ligand pK$_a$ as a function of the distance *z*, along the membrane normal, between the ligand terminal and the center of mass of the membrane. We set up a thermodynamic cycle, as previously reported by Mac Callum *et al.*[37] and shown in the first panel of Fig. 3. The two horizontal segments of the cycle correspond to the free energy of transfer of the ligand (protonated, $\Delta G^p_{w \to m}$ or deprotonated, $\Delta G^{dep}_{w \to m}$) from the water phase to distance *z* from the center of the membrane. The free energy profiles of the deprotonated and protonated ligands are shown in the central panel of Fig. 3, as obtained with the metadynamics simulations. The offset between the two free energy profiles (right vertical segment of the cycle in Fig. 3) is provided by the $pK_a$ of the ligand in the water phase, which we assume to be 6.3 as measured by Moglianetti *et al*[35]. The offset has been calculated via the Henderson-Hasselbalch equation:

$$pK_a = pH - \log_{10}\left(e^{\frac{-\Delta G^{dep \to p}}{k_B T}}\right) \quad (1)$$

The cycle can thus be used to calculate the unknown free energy change associated to ligand protonation at distance *z* from the center of the membrane, $\Delta G^m_{dep \to p}$, which is then related to the ligand $pK_a$ via equation (1). The right panel of Fig. 3 shows the resulting $pK_a$ as a function of *z*. The shaded area shows a conservative estimate of the indeterminacy with which the $pK_a$ of the ligand in water is known: the free ligand has a $pK_a$ of 5 (bottom edge of the shaded area for *z* = 3 nm from the membrane center), while cooperative effects on the NP surface have been predicted to shift the $pK_a$ of similar ligands up to 7.6[34] (upper edge of the shaded area). Using 6.3 as reference $pK_a$ value in water[35], the light-blue profile is obtained and the error bars deriving from our metadynamics calculations lay within the shaded area. The shaded grey areas indicate the distribution of lipid heads and glycerol groups. The $pK_a$ equals the physiological pH of 7.4 in the lipid heads region, and all the charged ligands reaching down to the glycerol region should be protonated and be able to translocate to the distal leaflet of the membrane without perturbing membrane structure or induce water transfer.

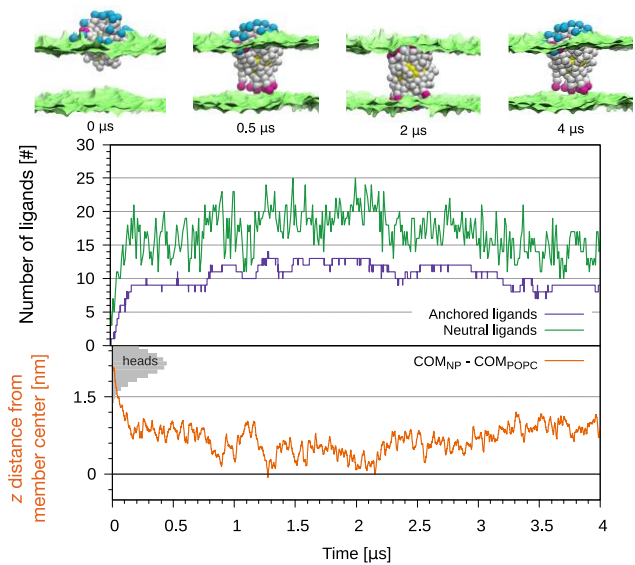

**Fig. 4**. Top: the figures show the configurations of the NP as it is progressively embedded in the membrane. Charged beads in blue, protonated beads in fuchsia and hydrophobic beads in white. Lipid heads are shown in green as surface representation, water and lipid tails are not shown. Bottom: the number of anchored ligands (violet), the protonated ligands (green) and $d_z$ for the NP center of mass as a function of the simulation time. The shaded grey areas indicate the distribution of lipid heads of the entrance leaflet.

We further exploited the knowledge of the $z$-dependent $pK_a$ of the titrable sites of the anionic ligands to perform constant-pH simulations of the NP-membrane interaction. One run was initialized as in the top left panel of Fig. 1 and, at regular intervals of time ($\Delta t$ = 10 ns), the protonation state of each ligand was reassigned based on its $pK_a$ value. The choice of $\Delta t$ is arbitrary and affects the kinetics of the process, but this setup allows to monitor membrane deformations, during the translocation of many neutral –COOH terminated beads, without the interference of any bias potential. During the run we observe that, as the number of protonated ligands increases, the translocation events increase too, as shown in the top graph of Fig. 4. Coherently, the NP penetrates deeper and deeper into the bilayer, as shown in the bottom graph of Fig. 4. If the anchored ligand remains protonated, the back-transition is favorable as well, causing some fluctuations on the number of anchored ligands. We remark that none of these anchoring and dis-anchoring events was accompanied by translocation of water beads, and we did not observe any significant membrane deformation during the anchoring process. The first anchoring event of a protonated ligand occur in few ns. Then, after, about 0.2 µs, the NP is stably inserted in the membrane. At this stage, the number of anchored ligands fluctuates around 10 while the number of protonated ligands fluctuates between 10 and 20. From about 1 µs on we observe that some ligands anchored to the distal leaflet start to change their protonation state, coming back to the negatively charged state and making the back-transition unfavorable. In fact, we do not observe any back-transition for the ligands that becomes negatively charged. From 1 to 2 µs the number of anchored ligands increases up to 15 and the NP distance from the membrane COM vanishes and becomes, in some cases, even negative. The NP thus results fully immersed in the membrane with roughly half ligands anchored to the entrance leaflet, and half to the distal one. This situation is subject to fluctuations – indeed, after ~ 2 µs the number of anchored ligands suddenly decreases and the NP gets back to a distance of ~ 0.5 nm from the membrane COM.

We can speculate that, due to fluctuations, the distance between the NP and the center of the bilayer could become more and more negative, leading to a complete transition of the NP from the entrance to the distal leaflet.

## Conclusions

Anionic Au NPs functionalized by MUA ligands can thus interact with zwitterionic lipid membranes via a mechanism that is common to charged amino acids and cell-penetrating peptides: similar $pK_a$ shifts have been reported for negatively charged amino acids with carboxylate groups[37] (Asp, Glu), and Ala-based pentapeptides[38] in different lipid environments[39]. The protonation of carboxylate groups is the key to the membrane insertion mechanism of pH (low) insertion peptides (pHLIP®)[40,41] as well. These peptides, which at pH 7.4 do not enter the membrane core, are designed to adopt a transmembrane helical configuration in presence of an acidic environment, as that of tumors. They have thus been exploited for tumor imaging and also have been shown to allow for the delivery of moderately hydrophilic drug cargos inside the diseased cells[42]. For pHLIPs, the transition from the membrane-adsorbed state to the transmembrane state is triggered by the protonation of 2-4 carboxyl groups[43,44]. Here we have shown that the interaction of the MUA ligands with the phosphocholine membrane is spontaneous also at physiological pH, due to the presence of a single carboxyl group in each ligand and to the rapid increase of its $pK_a$ in the region of the lipid headgroups. We thus envisage that, though not pH-selective, the non-disruptive interaction of MUA-functionalized NPs with plasma membranes at physiological pH could be exploited in a similar way as that of pHLIPs peptides for the delivery of hydrophilic cargos to the cell interior. In more general terms, we anticipate that the carboxyl-containing ligand protonation could be exploited for the design of NPs with a stable, controlled and less disruptive interaction with cell membranes.

## Conflicts of interest

There are no conflicts to declare.

## Acknowledgements

Giulia Rossi acknowledges funding from the ERC Starting Grant BioMNP – 677513. Calculations were in part carried out at CINECA (grant HP10CRSL8N to GR).

## Notes and references